\documentclass{aa}
\usepackage{txfonts}
\usepackage[dvips]{graphicx}
\def\I{\mbox{IRAS\,20126+4104}}

\def\WAT{H$_2$O}
\def\METH{CH$_3$OH}

\def\kms{\mbox{km~s$^{-1}$}}
\def\Jyb{\mbox{Jy~beam$^{-1}$}}
\def\mJyb{\mbox{mJy~beam$^{-1}$}}

\def\Vlsr{\mbox{$V_{\rm LSR}$}}

\begin{document}
%\thesaurus{08(02.13.3; 09.09.1; 09.10.1; 13.19.3)}
\title{
Methanol and water masers in \I:\\
The distance, the disk, and the jet\thanks{Based on
observations carried out with the NRAO Very Large Baseline Array and European VLBI Network.}
}
%\subtitle{...}
\author{L. Moscadelli \inst{1} \and R. Cesaroni \inst{1} \and M.J. Rioja \inst{2,3} \and R. Dodson \inst{2} \and M.J. Reid \inst{4}}
\institute{
	   Osservatorio Astrofisico di Arcetri, INAF, Largo E. Fermi 5,
           I-50125 Firenze, Italy \email{cesa@arcetri.astro.it}
\and
           International Centre for Radio Astronomy Research, University of Western Australia, Perth, Australia
\and
	   Observatorio Astron\'omico Nacional, Apdo. 112, E-28803 Alcal\'a de Henares, Madrid, Spain
\and
           Harvard-Smithsonian Center for Astrophysics, 60 Garden Street, Cambridge, MA 02138, USA
}
\offprints{L. Moscadelli, \email{mosca@arcetri.astro.it}}
\date{Received date; accepted date}

%\markboth{???? et al.: }{}
\titlerunning{Methanol and water masers in \I}
\authorrunning{L. Moscadelli et al.}

\abstract{
%Context
Knowledge of the distance to high-mass star forming regions is crucial to
obtain accurate luminosity and mass estimates of young OB-type (proto)stars
and thus better constrain their nature and age. \I\ is a
special case, being the best candidate of a high-mass (proto)star surrounded
by an accretion disk. Such a fact may be used to set constraints on theories
of high-mass star formation, but requires confirmation that the mass
and luminosity of \I\ are indeed typical of a B0.5 star, which in turn
requires an accurate estimate of the distance.
}{
%Aims
The goal of our study is twofold: to determine the distance to \I, using
the parallax of \WAT\ masers associated with the source, and unveil the
3D velocity field of the disk, through proper motion measurements of
the 6.7~GHz \METH\ masers. At the same time, we can also obtain an
estimate of the systemic velocity in the plane of the sky of the disk+star
system.
}{
%Methods
We used the Very Long Baseline Array and the European VLBI Network to
observe the 22.2~GHz \WAT\ and 6.7~GHz \METH\ masers in \I\ at a number
of epochs suitably distributed in time. The absolute positions of the maser
features
were established with respect to reference quasars, which allowed us
to derive absolute proper motions.
}{
%Results
From the parallax of the \WAT\ masers we obtain a distance of
$1.64\pm0.05$~kpc, which is very similar to the value adopted so far in the
literature (1.7~kpc) and confirms that \I\ is a high-mass (proto)star. From
the \METH\ masers we derive the component in the plane of the sky of the
systemic velocity of the disk+star system ($-16$~\kms\ in right-ascension and
+7.6~\kms\ in declination). Accurate knowledge of the distance and systemic
velocity allows us to improve on the model fit to the \WAT\ maser jet presented
in a previous study.
Finally, we identify two groups of \METH\ maser
features, one undergoing rotation in the disk and possibly distributed along
a narrow ring centered on the star, the other characterised by relative
proper motions indicating that the features are moving away from the disk,
perpendicular to it. We speculate that the latter group might be tracing the
disk material marginally entrained by the jet.
}{
%Conclusions
VLBI multi-epoch observations with phase referencing are confirmed to be
an excellent tool for distance determinations and for the investigation
of the structure and 3D velocity field within a few 100~AU from newly
born high-mass stars.
}
\keywords{Stars: formation -- ISM: individual objects: \I\ -- Masers -- Proper motions}

\maketitle

\section{Introduction}

The role of disks in the formation of high-mass (i.e. O- and early B-type) stars
has been extensively discussed in a number of reviews (see e.g.
Cesaroni et al.~\cite{natu}, \cite{ppv}). Disks are important because they
focus the accretion onto the forming star
and allow some of the stellar photons to escape along the disk axis
(see Krumholz et al.~\cite{krum} and references therein).
The combined effects of these two mechanisms might prevent radiation pressure
from halting the infalling gas, thus allowing growth of the stellar mass beyond
the otherwise maximum value of $\sim8~M_\odot$ (Palla \& Stahler~\cite{past}).
For these reasons, the detection of disks around early-type (proto)stars
could be used to prove that the formation of such stars is a scaled up version
of that of solar-type stars.

Despite the numerous searches made in recent years with (sub)millimeter
interferometers, no true (i.e. self-gravitating, dynamically stable)
circumstellar disk has been found associated with early O-type (proto)stars.
There is, however, growing evidence of Keplerian disks around early B-type
stars and \I\ is an exemplary case. This $10^4~L_\odot$
young stellar object (YSO) has been the subject of a long series of studies,
with angular resolution spanning from a few 10\arcsec\ to a few milli-arcsec
(mas).  Here, we note some findings most relevant to the present article,
namely:
\begin{itemize}
\item \I\ is associated with a Keplerian disk rotating about a $\sim7~M_\odot$
 protostar (Cesaroni et al. \cite{cesa97}, \cite{cesa99}, \cite{cesa05};
 Zhang et al. \cite{zhang}).
\item The disk is also seen in silhouette as a dark lane at near- and mid-IR
 wavelengths (Sridharan et al.~\cite{srid}; De Buizer et al.~\cite{debu}).
\item The YSO is powering a bipolar jet/outflow, imaged in a variety
 of tracers from $\sim$1\arcmin\ down to $\sim$100~mas from the star
 (Wilking et al.~\cite{wilk};
 Cesaroni et al. \cite{cesa97}, \cite{cesa99}, \cite{cesa05};
 Shepherd et al.~\cite{shep}; Su et al.~\cite{su}; Shinnaga et al.~\cite{shin};
 Hofner et al.~\cite{hof99}, \cite{hof07};
 Moscadelli et al.~\cite{mosca00}, \cite{mosca05};
 Trinidad et al.~\cite{trini}).
\item The jet/outflow appears to be undergoing precession
 (Shepherd et al.~\cite{shep}; Cesaroni et al.~\cite{cesa05}; Lebr\'on et
 al.~\cite{lebr}), perhaps caused by the presence of another YSO detected at
 radio and IR wavelengths (Hofner et al.~\cite{hof99}; Sridharan et
 al.~\cite{srid}).
\end{itemize}

Such an overwhelming quantity of information made it possible to determine
a number of physical parameters of the star-disk-outflow system.
Among these, the most important are the stellar mass and luminosity,
which allows one to identify this object as an early B-type protostar.
However, the values on which this conclusion is based depend on the
distance, $d$, which is poorly known. The value quoted in the literature
(1.7~kpc) was adopted under the assumption that \I\ lies in the
Cyg-X region (see Shinnaga et al.~\cite{shin} for a discussion of the
distance determination). In a previous study, we tried to constrain the
distance to the source using the observed proper motions of the \WAT\ maser
spots associated with it (see Moscadelli et al.~\cite{mosca05} for details).
We concluded that \I\ should be located farther than 1.2~kpc, but this result,
beside being a very loose constraint, is also model dependent. In conclusion,
one cannot exclude the possibility that this YSO is much less massive than
expected, if $d\ll1.7$~kpc, or much more, if $d\gg1.7$~kpc. The former
possibility would dramatically reduce the values of stellar mass and
luminosity, thus ``destroying'' the most convincing example of an accretion
disk around an early B-type protostar. In turn, this would have implications
for the formation scenario of high-mass stars, as previously explained. Hence,
it is of fundamental importance to obtain an accurate estimate of $d$.
Recently, several studies (see Reid et al.~\cite{reid09b} and references
therein) have proved the potential of maser parallax measurements for
distance determinations. We have applied this method to the \WAT\ masers in
\I\ and in Sect.~\ref{spara} we report on the results obtained.

Another crucial issue in Galactic studies is that of the reference system
used to determine absolute proper motions. In the case of \I, it is necessary
to refer the measurement to the star. Usually, for this purpose the absolute
proper motion is measured by referring the spot position to a background
quasar. Then the apparent motions due to the annual parallax and motion of
the source with respect to the Sun are subtracted. The result, however, is
not very precise, as the corrections depend on the distance to the source and
the accuracy of the Solar motion and Galactic rotation curve adopted.  In the
case of \I, the distance can be determined from the parallax measurement of
the \WAT\ maser spots (as shown in the present paper) and the major source of
uncertainty remains the reliability of the Galactic rotation curve, for which
deviations of $\sim$15~\kms\ are known to exist (see Reid et
al.~\cite{reid09b}). This problem would be solved if one could directly
measure the proper motion of the star and then subtract this from that of the
maser spots. In Sect.~\ref{svsys} we will show that the \METH\ masers in
\I\ can be used for this purpose.

Finally, in Sect.~\ref{sdj} knowledge of the distance and 3D velocities
of the \WAT\ and \METH\ maser spots will be used to analyse the detailed
structure of the disk-jet system in \I\ on scales as small as $\sim$150~AU
from the protostar.

\section{Observations and calibration}
\label{sobs}

We conducted multi-epoch, VLBI observations of the H$_2$O and CH$_3$OH masers
toward \I\ at 22.2 and 6.7~GHz, respectively. To determine the
maser absolute positions and velocities, we performed phase-referencing
observations by fast switching between the maser source and the calibrator
J2007$+$4029. This calibrator has an angular offset from the maser source of
$1\fdg5$ and belongs to the list of sources defining the International
Celestial Reference Frame (ICRF). Its absolute position is known to better
than $\pm1$~mas and its flux measured with the
NRAO\footnote{The National Radio Astronomy Observatory is operated by
Associated Universities, Inc., under cooperative agreement with the National
Science Foundation.}
Very Long Baseline Array (VLBA) at S and X bands is 0.3 and 0.8~\Jyb,
respectively.  At each VLBI epoch, H$_2$O and CH$_3$OH maser absolute
positions have been derived by imaging the maser visibilities after applying
the phase and amplitude corrections derived working with the strong
calibrator J2007$+$4029. The estimated accuracy of the maser position
offsets  is \ 0.05--0.1~mas  and \ $\sim$0.3~mas for the H$_2$O and CH$_3$OH
masers, respectively, the better accuracy of the water maser positions owing
to the improved calibration of the delays
introduced  by the Earth's atmosphere (see Section~\ref{vlba_water}).

Six fringe finders (J0237+2848; 3C345; J1800+3848; J2002+4725; J2005+7752;
3C454.3) were observed for bandpass, single-band delay, and instrumental
phase-offset calibration.  Data were reduced with AIPS following the VLBI
spectral line procedures.  For a description of the general data calibration
procedures and the criteria used to identify individual masing clouds, derive
their parameters (position, intensity, flux and size), and measure their
(relative and absolute) proper motions, we refer to the recent paper on VLBI
observations of H$_2$O and CH$_3$OH masers by Sanna et al.~(\cite{sanna}).
For a description of the general observational setup and method of
calibration of maser parallax observations, we refer to Reid et
al.~(\cite{reid09a}).  The derived absolute proper motions have been
corrected for the apparent proper motion due to the Earth's orbit around
the Sun (parallax), the Solar Motion and the differential Galactic Rotation
between our LSR and that of the maser source. We have adopted a flat Galaxy
rotation curve, with a standard Solar motion of 19.5~\kms\ towards the Solar Apex at
$\alpha({\rm J2000})=271\degr$ and $\delta({\rm J2000})=29\degr$, corresponding
to U,V,W=-10.4,14.8,7.3~\kms\ (see Mihalas \& Binney~\cite{mibi}), and the
values of the Galactic Constants ($R_0$ = 8.4~kpc, $\Theta_0$ = 254~\kms) as
recently determined by Reid et al.~(\cite{reid09b}).  In the following we
only report on the main observational parameters of our maser VLBI
observations.

\subsection{22.2~GHz H$_2$O masers: VLBA parallax observations}
\label{vlba_water}

We observed \I\ (tracking center: $\alpha(J2000)=20^h14^m26\fs0253$
and $\delta(J2000)=41\degr13\arcmin32\farcs666$) with the VLBA in the
$6_{16}$--$5_{23}$ H$_2$O transition (rest frequency 22.23507985~GHz). The
observations (program code: BM269) consisted of 5 epochs: January 28, May
24, August 2, November 5 2008, and January 26, 2009. These dates were
selected to symmetrically sample both the eastward and northward parallax
signatures and minimize correlations among the parallax and proper motion
parameters.  At each epoch, in two 4~h blocks of phase-referencing
observations (using a switching cycle between calibrators of 1~min),
we recorded both circular polarizations using two 8~MHz
bandwidths, with the higher-frequency bandwidth centered on a LSR velocity of
--3.0~\kms.  Geodetic-style observing blocks of 40~min were placed
before the start, in the middle, and after the end of our phase-reference
observations, in order to monitor slow changes in the total atmospheric delay
for each telescope.  The data were processed with the VLBA FX correlator in
two correlation passes using either 512 or 16 spectral channels for the maser
line and calibrator data, respectively. In each correlator pass, the data
averaging time was 1~s.

The natural CLEAN beam was an elliptical Gaussian with a FWHM size of about $0.7~\textrm{mas} \times 0.4~\textrm{mas}$ at a P.A. of about $-15\degr$ (east of north), 
slightly varying from epoch to epoch. To better compare maser images across the epochs, maser maps have been reconstructed using a round beam with a FWHM size of 1~mas. 
Using an on-source integration time of about 4~h, the rms noise level on the channel maps 
(evaluated over an area where no signal is present)  was $\sim$5~\mJyb. 
With 512 correlator channels, the 8~MHz band provides a resolution in the line-of-sight velocity of 0.2~\kms.

\subsection{22.2~GHz H$_2$O masers: single-epoch EVN observation}

We observed \I\ (tracking center: $\alpha(J2000)=20^h14^m26\fs0387$ and
$\delta(J2000)=41\degr13\arcmin32\farcs534$) with the European VLBI
Network\footnote{The European VLBI Network is a joint facility of European,
Chinese and other radio astronomy institutes founded by their national
research councils.} (EVN) in the H$_2$O 22.2~GHz maser line on March 4 2009
(program code: EM059D).  The antennas involved in the observations were
Jodrell2, Effelsberg, Medicina, Noto, Metsahovi, Onsala, Yebes and Shanghai.
During a total run of 13~h (interspersing phase-referencing, maser-calibrator
observations every 1.5~h with scans on the fringe-finders), we recorded both
circular polarizations using a bandwidth of 16~MHz, centered on a LSR
velocity of --3.5~\kms.  The data were processed with the MKIV correlator at
the Joint Institute for VLBI in Europe (JIVE -- Dwingeloo, The Netherlands)
using an averaging time of 2~s and 1024 spectral channels for each observing
bandwidth.

The natural CLEAN beam was an elliptical Gaussian with a FWHM size of
1.5~mas$\times$1.1~mas at a P.A. of 83\degr. With an
on-source integration time of about 4~h, the rms noise level on the channel
maps was about 30~\mJyb.  With 1024 correlator channels, the 16~MHz band
provides a resolution in the line-of-sight velocity of 0.2~\kms.

\begin{table*}
\begin{flushleft}
\caption[]{Parameters of the 6.7~GHz \METH\ maser features in \I.
 The offsets, $\Delta\alpha$ and $\Delta\delta$, of a given feature are
 measured at the epoch (given in column~2) when the feature was
 detected for the first time, and are computed with respect to the position of feature~3 at the
 same epoch.  The positions of feature~3 at the three observing epochs are:
 $\alpha({\rm J2000}) = 20^{\rm h} 14^{\rm m} 26\fs07111$,
 $\delta({\rm J2000}) = 41\degr 13' 32\farcs701$, at epoch 1 (November 6, 2004);
 $\alpha({\rm J2000}) = 20^{\rm h} 14^{\rm m} 26\fs07026$,
 $\delta({\rm J2000}) = 41\degr 13' 32\farcs691$, at epoch 2 (March 21, 2007);
 $\alpha({\rm J2000}) = 20^{\rm h} 14^{\rm m} 26\fs06963$,
 $\delta({\rm J2000}) = 41\degr 13' 32\farcs683$, at epoch 3 (March 11, 2009).
 The errors on these positions are in all cases $<$1~mas.
 The relative proper motions, $v_{\rm \alpha}$ and $v_{\rm \delta}$ also refer
 to feature~3.
 }
\label{tmeth}
\begin{tabular}{cccccccccc}
\hline
\hline
feature & epochs of & $\Delta\alpha$ & $\Delta\delta$ & $S_\nu$ & $V_{\rm LSR}$ & $v_{\rm \alpha}$ & $v_{\rm \delta}$ & $V_{\rm \alpha}$ & $V_{\rm \delta}$ \\
number & detection & (mas) & (mas) & (Jy) & (\kms) & (\kms) &(\kms) & (\kms) &(\kms)  \\
\hline
 1 & 1,2,3 & $-$233.00$\pm$0.08 &   $-$5.79$\pm$0.08 & 40.5 & $-$6.1 &    $-$2.9$\pm$0.2 &     0.2$\pm$0.2 	 & 	 $-$16.6$\pm$0.8	 & 	   6.4$\pm$0.8	 \\  
 2 & 1,2,3 & $-$40.54$\pm$0.09 &   31.24$\pm$0.09 &  5.8 & $-$7.7 &    $-$2.4$\pm$0.2 &     0.9$\pm$0.2 	 & 	 $-$16.1$\pm$0.8	 & 	   7.2$\pm$0.8	 \\  
 3 &  1,2,3 &  0.00 &    0.00  &  4.9 & $-$7.1 &  0.0  &  0.0  	 & 	 $-$13.6$\pm$0.8	 & 	   6.2$\pm$0.8	 \\  
 4 & 1,2,3 & $-$171.13$\pm$0.10 &  $-$19.35$\pm$0.10 &  3.7 & $-$6.5 &    $-$4.8$\pm$0.2 &     3.5$\pm$0.2 	 & 	 $-$18.4$\pm$0.8	 & 	   9.6$\pm$0.8	 \\  
 5 & 1,3 &$-$23.75$\pm$0.30 &   27.79$\pm$0.29 &  3.4 & $-$7.6 &  --  &  --  	 & 	   -- 	 & 	   -- 	 \\  
 6 &  3 &$-$3.28$\pm$0.08 &   $-$0.11$\pm$0.09 &  2.8 & $-$6.8 &  --  &  --  	 & 	   -- 	 & 	   -- 	 \\  
 7 & 1,3 &$-$31.45$\pm$0.30 &   27.94$\pm$0.33 &  2.6 & $-$7.6 &  --  &  --  	 & 	   -- 	 & 	   -- 	 \\  
 8 & 1,2,3 & $-$152.15$\pm$0.20 &  $-$17.98$\pm$0.21 &  2.5 & $-$6.7 &    $-$3.5$\pm$0.3 &     1.5$\pm$0.3 	 & 	 $-$15.7$\pm$0.8	 & 	   8.1$\pm$0.8	 \\  
 9 & 1,2,3 & $-$246.24$\pm$0.26 &   $-$1.89$\pm$0.24 &  2.5 & $-$6.8 &    $-$4.9$\pm$0.4 &     1.8$\pm$0.4 	 & 	 $-$16.7$\pm$0.9	 & 	   7.8$\pm$0.9	 \\  
10 &  1,2,3 & 43.46$\pm$0.10 &   $-$1.48$\pm$0.10 &  1.7 & $-$8.3 &    $-$1.3$\pm$0.2 &    $-$0.4$\pm$0.2 	 & 	 $-$14.9$\pm$0.8	 & 	   5.9$\pm$0.8	 \\  
11 &  1,2,3 & $-$79.31$\pm$0.10 & $-$113.86$\pm$0.10 &  1.1 & $-$4.9 &    $-$0.4$\pm$0.2 &     5.4$\pm$0.2 	 & 	 $-$14.1$\pm$0.8	 & 	  11.7$\pm$0.8	 \\  
12 &  1,2 & $-$16.71$\pm$0.16 &    3.96$\pm$0.17 &  1.0 & $-$6.9 &    --  &     --	 & 	 --	 & 	   --	 \\  
13 &  3 & $-$11.51$\pm$0.25 &  $-$34.93$\pm$0.16 &  0.9 & $-$7.6 &  --  &  --  	 & 	   -- 	 & 	   -- 	 \\  
14 & 2,3 & $-$215.05$\pm$0.23 &  $-$12.07$\pm$0.19 &  0.9 & $-$6.3 &  --  &  --  	 & 	   -- 	 & 	   -- 	 \\  
15 &  2,3 & $-$30.22$\pm$0.11 &    2.63$\pm$0.11 &  0.9 & $-$7.1 &  --  &  --  	 & 	   -- 	 & 	   -- 	 \\  
16 & 1,2,3 & $-$176.95$\pm$0.17 &  $-$29.11$\pm$0.17 &  0.9 & $-$5.6 &    $-$2.8$\pm$0.3 &     1.3$\pm$0.3 	 & 	 $-$16.0$\pm$0.8	 & 	   7.9$\pm$0.8	 \\  
17 & 1,2,3 & $-$180.45$\pm$0.11 &  $-$52.37$\pm$0.12 &  0.8 & $-$5.1 &    $-$2.9$\pm$0.3 &    $-$1.1$\pm$0.3 	 & 	 $-$16.4$\pm$0.8	 & 	   5.1$\pm$0.8	 \\  
18 & 3 & $-$77.79$\pm$0.15 &    6.46$\pm$0.14 &  0.7 & $-$6.7 &  --  &  --  	 & 	   -- 	 & 	   -- 	 \\  
19 & 3 & $-$150.64$\pm$0.25 &  $-$80.58$\pm$0.21 &  0.6 & $-$6.6 &  --  &  --  	 & 	   -- 	 & 	   -- 	 \\  
20 &  1,2 & 27.36$\pm$0.23 &   $-$0.98$\pm$0.23 &  0.5 & $-$7.8 &   --  &     -- 	 & 	 --	 & 	  --	 \\  
21 & 3 & $-$140.40$\pm$0.15 &  $-$12.30$\pm$0.12 &  0.4 & $-$5.5 &  --  &  --  	 & 	   -- 	 & 	   -- 	 \\  
22 & 3 & $-$214.16$\pm$0.21 &   $-$5.22$\pm$0.17 &  0.4 & $-$6.6 &  --  &  --  	 & 	   -- 	 & 	   -- 	 \\  
23 & 3 & $-$152.48$\pm$0.13 &  $-$11.28$\pm$0.14 &  0.4 & $-$5.6 &  --  &  --  	 & 	   -- 	 & 	   -- 	 \\  
24 &   1,2 & 7.47$\pm$0.38 &   $-$3.15$\pm$0.40 &  0.3 & $-$6.8 &  --  &  --  	 & 	   -- 	 & 	   -- 	 \\  
25 &  2,3 & $-$189.83$\pm$0.44 &  $-$22.76$\pm$0.28 &  0.2 & $-$5.8 &  --  &  --  	 & 	   -- 	 & 	   -- 	 \\  
26 &  2,3 & $-$32.69$\pm$0.85 &  $-$75.49$\pm$0.65 &  0.2 & $-$5.1 &  --  &  --  	 & 	   -- 	 & 	   -- 	 \\  
27 &  3 &  46.71$\pm$0.15 &   29.06$\pm$0.20 &  0.2 & $-$8.7 &  --  &  --  	 & 	   -- 	 & 	   -- 	 \\  
28 &  3 & 36.86$\pm$0.14 &    4.74$\pm$0.15 &  0.2 & $-$8.3 &  --  &  --  	 & 	   -- 	 & 	   -- 	 \\  
29 &  2,3 & $-$212.65$\pm$0.36 &  $-$16.60$\pm$0.36 &  0.1 & $-$5.7 &  --  &  --  	 & 	   -- 	 & 	   -- 	 \\  
30 & 3 & $-$191.04$\pm$0.20 &  $-$33.16$\pm$0.22 &  0.1 & $-$5.2 &  --  &  --  	 & 	   -- 	 & 	   -- 	 \\  
31 &  3 & $-$228.56$\pm$0.38 &  $-$19.26$\pm$0.40 &  0.1 & $-$4.7 &  --  &  --  	 & 	   -- 	 & 	   -- 	 \\  
32 &  2,3 & $-$67.77$\pm$0.76 &  $-$84.08$\pm$0.82 &  0.1 & $-$5.2 &  --  &  --  	 & 	   -- 	 & 	   -- 	 \\  
33 &  3 & $-$44.31$\pm$0.51 &  $-$72.72$\pm$0.40 &  0.1 & $-$4.6 &  --  &  --  	 & 	   -- 	 & 	   -- 	 \\  
\hline
%________________________________________________________________________________________________________________________
%Absolute Positions (J2000)
%
%1) : $\alpha(J2000) = 20^{\rm h} 14^{\rm m} 26\fs07111$ ,  $\delta(J2000) = 41\degr 13' 32\farcs701$
%
%2) : $\alpha(J2000) = 20^{\rm h} 14^{\rm m} 26\fs07026$ ,  $\delta(J2000) = 41\degr 13' 32\farcs691$
%
%3) : $\alpha(J2000) = 20^{\rm h} 14^{\rm m} 26\fs06963$ ,  $\delta(J2000) = 41\degr 13' 32\farcs683$
\end{tabular}

\end{flushleft}
\end{table*}

\subsection{6.7~GHz CH$_3$OH masers: multi-epoch EVN observations}

We observed \I\ (tracking center: $\alpha(J2000)=20^h14^m26\fs05$ and
$\delta(J2000)=41\degr13\arcmin32\farcs7$) with the European VLBI Network (EVN) in
the $5_{1}-6_{0} A^+$ CH$_3$OH transition (rest frequency 6.66851920~GHz), at
two epochs (program codes: EM064C and EM064D), separated by about 2~yr,
observed on March 21, 2007 and March 11, 2009.  At both epochs, antennas
involved in the observations were Cambridge, Jodrell2, Effelsberg, Onsala,
Medicina, Noto, Torun and Westerbork.  During a run of about 6~h per epoch
(alternating phase-referencing, maser-calibrator observations every 1.5~h
with scans on the fringe-finders), we recorded both circular polarizations
using eight 2~MHz bandwidths, with the third (frequency-ordered) band
centered on the maser target LSR velocity of \ $-$6.1~\kms.  The total 16~MHz
bandwidth for each polarization was effective to increase the SNR on the
continuum calibrator.  The data were processed with the MKIV correlator at
the Joint Institute for VLBI in Europe (JIVE -- Dwingeloo, The Netherlands) in
two correlation passes using either 1024 or 128 spectral channels for the
maser line and calibrator data, respectively. In each correlator pass, the
data averaging time was 1~s.

The natural CLEAN beam was an elliptical Gaussian with a FWHM size of about $8~\textrm{mas} \times 6~\textrm{mas}$ at a P.A. of about \ $\sim45\degr$,
slightly varying between the two epochs. 
Using an on-source integration time of about 2.2~h, the rms noise level on the channel maps was about \ 6~\mJyb. 
With 1024 correlator channels, the 2~MHz band provides a resolution in the line-of-sight velocity of 0.09~\kms.

In order to unambiguously determine maser spot persistency and derive
reliable proper motions, it is fundamental to observe (at least) three
different epochs (see discussion in Sanna et al.~\cite{sanna}). For this
purpose, we have extracted from the EVN archive and analysed
phase-referencing observations of the  6.7~GHz CH$_3$OH masers towards
\I\ observed on November 6 2004 (program code: EL032). The main parameters of
these observations closely match those of our two EVN epochs. The selected
calibrator, J2007$+$4029, is the same intense quasar.  The EVN
antennas used are the same as in the programs EM064C and EM064D, with the
exception that the Darnhall telescope was used instead of Jodrell2. For the
maser emission, the observing bandwidth, 2~MHz, and the correlator spectral
channels, 1024, are the same as employed in our, more recent, EVN
experiments.  As a result, the angular and line-of-sight velocity resolution,
and the sensitivity of the EL032 data are comparable to those of the EM064C
and EM064D experiments, and the three EVN observations constitute a suitable
dataset to study the maser motions.

\begin{figure}
\centering
\resizebox{8.5cm}{!}{\includegraphics[angle=0]{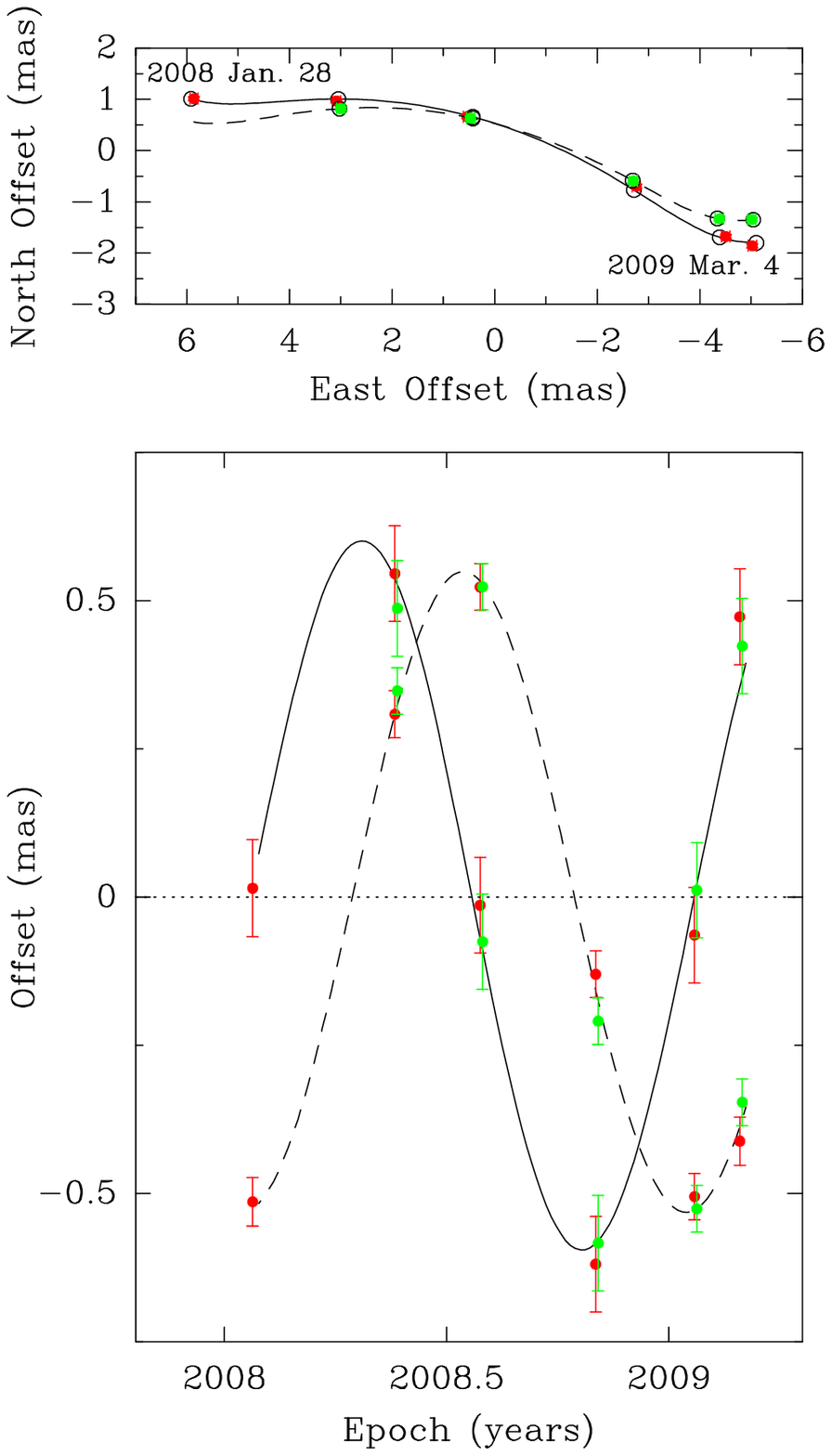}}
\caption{
{\bf Top.} Positions of the two persistent \WAT\ maser features in \I\ at
the six epochs of our VLBI observations. Note that the reported positions 
have been suitably offset from the measured positions for displaying purposes. 
Green and red circles indicate features
at --6.4 and --3.2~\kms, respectively.  
Their positions at the 2$^{\rm nd}$ observing epoch are
$\alpha(J2000)=20^h14^m26\fs0165$ and $\delta(J2000)=41\degr13\arcmin32\farcs577$,
for the feature at --3.2~\kms, and
$\alpha(J2000)=20^h14^m26\fs0167$ and $\delta(J2000)=41\degr13\arcmin32\farcs576$,
for the other.
The solid and dashed curves with open
circles are the best
fits, consisting of the apparent motion due to the annual parallax for
a common
a distance of $1.64\pm0.05$~kpc
for both features
plus a constant velocity in the plane of the sky
different for each feature (see text).
{\bf Bottom.} Residual motions in right ascension and declination, obtained after
subtraction of the constant velocities. The curves are the fits consisting
only of the annual parallax. The solid and dashed curves correspond,
respectively, to the R.A. and Dec. offsets.
}
\label{fwpm}
\end{figure}

\section{Results}

While many  \WAT\ maser features have been detected at the different epochs,
only two of them persisted long enough (over 5 and 6 epochs) to be used for
our parallax measurements. Details of the method adopted to derive a distance
estimate will be given in Sect.~\ref{spara}).  Since in the present article
our interest is focused on \WAT\ masers as tools to determine the distance to
\I, we will not report on the results obtained for the other shorter-lived
water maser features.

As for methanol masers, Table~\ref{tmeth} gives the main properties of the
detected \METH\ maser features. In particular, column~1 gives the feature
identification number, column~2
the observing epochs
at which the
feature was detected, columns~3 and~4 the position offsets with the
associated errors along the R.A. and Dec. axes, relative to the reference
feature~3, column~4 the flux density, column~5 the LSR velocity,
columns~6~and~7 the relative proper motion components with the associated
errors along the R.A. and Dec. axes, measured with respect to the reference
feature~3, and columns~8~and~9 the absolute proper motion components with
the associated errors along the R.A. and Dec. axes. Note that the positions
given in columns~3
and~4 are measured at the first of the epochs reported
in column~2,
while the quantities given in columns~5 and~6 are the mean values
over all the observing epochs.
The absolute
positions of feature~3 at the three observing epochs are given in the
caption of Table~\ref{tmeth}.
We stress that proper motions evaluated using only two epochs are affected by
much larger uncertainties (with errors from 0.6 up to 3~\kms), than proper
motions calculated using three epochs (typical errors are 0.3~\kms). The
uncertainty is mainly due to random variations in the spatial and velocity
structure of the maser emission. One needs at least three epochs to average
out this effect. Therefore, all the reported methanol maser proper motions
have been calculated using three epochs.

For the sake of clarity, it is important to point out
what is meant with ``relative'' and ``absolute'' in the present study.
When used for {\it positions}, the term relative refers to measurements
with respect to a maser feature, while absolute indicates
measurements
with respect to an external reference (the background quasar), whose
position is known to a high degree of accuracy. In the case of {\it proper
motions}, relative means that the proper motion is measured using a maser
feature
as a reference,
whereas absolute indicates proper motions obtained
by referring the motion to the background quasar, and then subtracting
the contribution of the parallax, Solar motion with respect to the LSR,
and galactic rotation. This is made in an attempt to obtain the proper motion,
and thus the velocity, relative to the protostar. However, the absolute
proper motions thus obtained require an additional correction, for which we
refer to the discussion in Sect.~\ref{svsys}.

\section{The distance to \I}
\label{spara}

Targets of our parallax measurements are compact water maser spots persistent
across the six VLBI (five VLBA and one EVN) epochs. Looking at the maser
spectra of each epoch, persistent emission could be identified only near the
systemic velocity at --3.5~\kms.  Inspecting the images of each velocity
channel, only two spots (at line-of-sight velocities of $-$6.4 and
$-$3.2~\kms, respectively) maintaned a
sufficiently
compact structure to be
used as astrometric targets. Such a low number of usable spots is not
surprising, due to the well known variability of \WAT\ maser emission (see
e.g. Felli et al.~\cite{felli}). Moreover, this emission happened to be
especially faint in \I\ during the period of our observations.

The results
of the multi-epoch \WAT\ measurements over $\sim$1.5~yr are illustrated in
Fig.~\ref{fwpm}.  The top panel shows the positions of the two persistent
maser features, one detected at all 6 epochs, the other only at 5 epochs.
These have been fitted with a model
taking into account the apparent motion due to the annual parallax plus a
constant velocity vector: --10.3~mas~yr$^{-1}$ in R.A. and
--2.7~mas~yr$^{-1}$ in Dec. for the feature at --3.2~\kms, and
--10.2~mas~yr$^{-1}$ in R.A. and --1.9~mas~yr$^{-1}$ in Dec. for the feature
at --6.4~\kms.  This vector accounts for both the motion of the local system
(the star) relative to the Sun and the motion of the spots relative to the
star. The assumption of constant velocity has proved successful in similar
experiments (see Reid et al.~\cite{reid09b} and references therein) and in
the case of \I\ is supported by the conical jet model of Moscadelli et
al. (\cite{mosca00}, \cite{mosca05}). The curves in Fig.~\ref{fwpm} are the
best fit to the data, corresponding to a parallax of $0\farcs61\pm0\farcs02$,
i.e. a distance of $1.64\pm0.05$~kpc. In the bottom panel of the figure, we
show the residual position offsets at the six epochs, after subtracting the
contribution of the constant velocity vector.

The fundamental conclusion that can be drawn from this result is that the
distance adopted so far in the literature (1.7~kpc) for \I\ was correct. This
confirms all the estimates of important physical parameters, in
particular the stellar mass ($\sim7~M_\odot$) and bolometric luminosity
($\sim10^4~L_\odot$). We can thus conclude that \I\ is indeed a B0.5
protostar surrounded by a Keplerian accretion disk
(Cesaroni et al.~\cite{cesa05}).

\section{The systemic velocity of \I}
\label{svsys}

\begin{figure*}
\centering
\resizebox{18cm}{!}{\includegraphics[angle=-90]{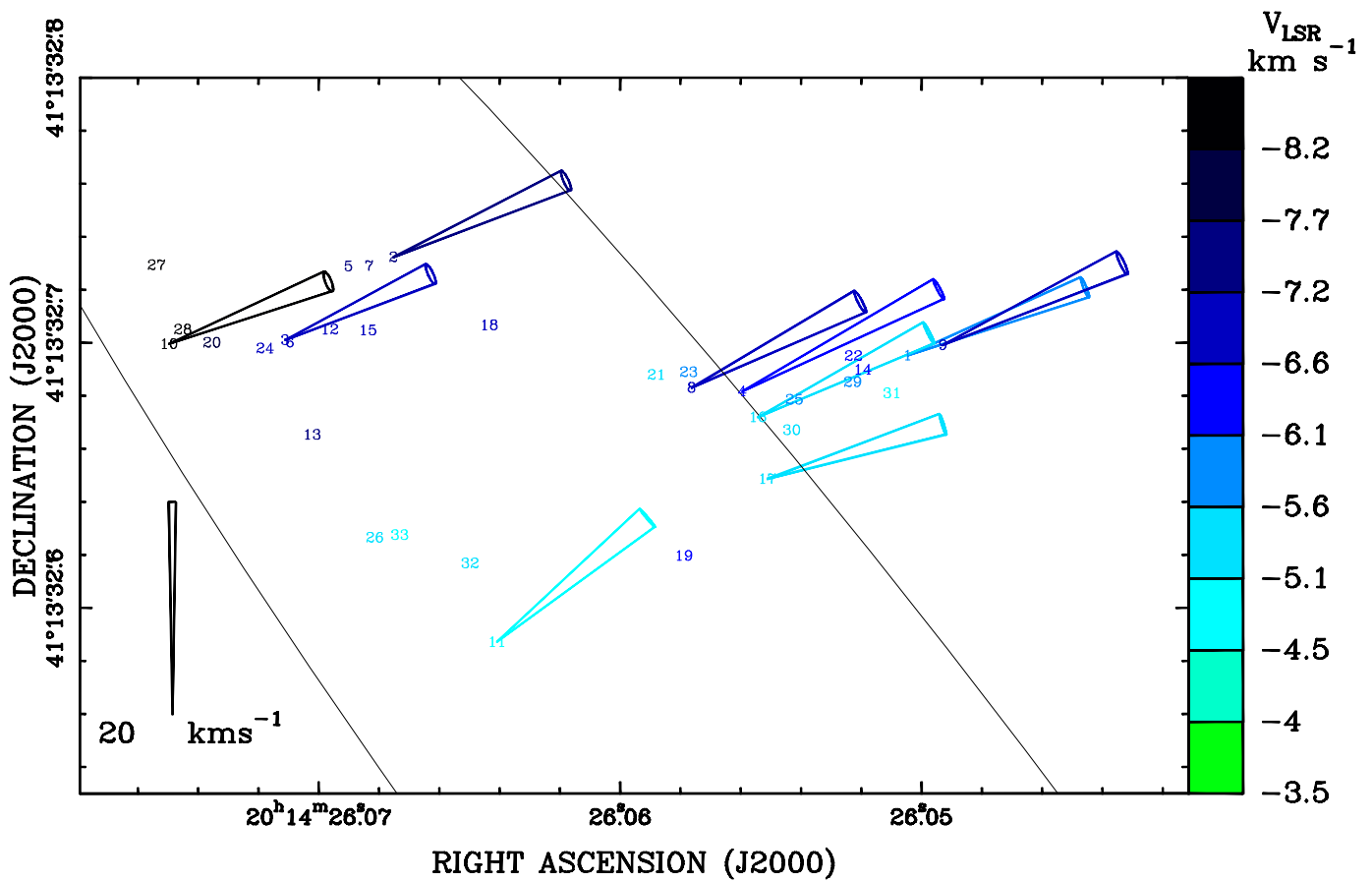}}
\caption{
Absolute proper motions (cones) of the \METH\ maser features (solid circles)
in \I.
Points without an associated cone have been detected only over one or
two epochs and the associated proper motion cannot be computed or is considered
unreliable.
Note that the all features are displayed at the positions of
the first epoch (November 6, 2004). For features undetected at this
epoch, the positions were obtained by extrapolation assuming they are
moving with the mean absolute proper motion of the other features
(in all cases the correction is $<$10~mas).
Absolute proper motions have been corrected for parallax, motion of the
Sun with respect to the LSR, and Galactic rotation
adopting a flat rotation curve as explained in Sect.~\ref{sobs}.
Most of the mean residual
proper motion ($\sim$18~\kms) is believed to be due to deviation
of the source from the assumed rotation curve.
The maser features are identified by the numbers given in column~1 of
Table~\ref{tmeth}, while their colour indicates the LSR velocity according
to the scale displayed to the right.
The two arcs schematically outline the Keplerian disk
(Cesaroni et al.~\cite{cesa05}),
whose center lies out of the figure, to the SW.
}
\label{fmpm}
\end{figure*}

Methanol masers have been claimed by several authors to be associated with
circumstellar disks (see e.g. Norris et al.~\cite{norris}, Minier et
al.~\cite{mini00}, Pestalozzi et al.~\cite{pesta}). In \I\ they appear
to lie in the north-eastern part of the disk (Minier et al.~\cite{mini01};
Edris et al.~\cite{edris}) and their velocities are blue-shifted, consistent
with the Keplerian rotation pattern observed in thermal lines. Our EVN images
are consistent with this finding, as illustrated in Fig.~\ref{fmpm}, where the
\METH\ maser features are overlayed on the disk silouhette. The latter is
very schematic and assumes a disk radius of $\sim$1000~AU (Cesaroni et
al.~\cite{ppv}) and the same center and inclination as the \WAT\ maser jet
(see Sect.~\ref{scone}).

If the masers lie in the disk and co-rotate with it, the corresponding
velocity vectors should also lie in the plane of the disk. Since the latter is
basically edge-on and the masers cluster is at a (projected) distance of
$\sim0\farcs2$ or 340~AU from the star, one would expect the proper motions
to be relatively small ($\la4$~\kms, assuming Keplerian rotation about a
7~$M_\odot$ star) and directed parallel to the disk, i.e. NE--SW. In
contrast, the absolute proper motions measured by us have speeds of
$\sim$18~\kms\ and are all roughly perpendicular to the disk, as shown in
Fig.~\ref{fmpm}. It is also worth noting that the dispersions of the
\METH\ velocities both along the line of sight and in the plane of the sky are
small compared to the mean speed of the features, unlike \WAT\ masers whose
velocities span an order of magnitude. Thus, we note that the
\METH\ masers are moving much slower than the \WAT\ masers.

Our explanation of all these facts is that {\it the absolute proper motions
of the \METH\ masers are dominated by a peculiar motion of about 10~\kms\
with respect to a circular Galactic orbit}. Whether such a
peculiar motion is due to a wrong estimate of the Solar motion with respect
to the LSR, to an inadequate Galaxy rotation curve, or to a deviation from it,
is irrelevant for our purposes. The important point is that the {\it
mean} absolute proper motion of the \METH\ masers can be used to correct any
absolute proper motion measurement and determine the true motion with respect
to the disk+star system to within about $\pm4$~\kms\ (the expected rotational
velocity). Incidentally, we note that deviations from the
standard Galaxy rotation curve are quite common, as demonstrated by Reid et
al. (\cite{reid09b}) in their study of the Galactic velocity field. Note
also that the existence of this ``spurious'' velocity component of
$\sim$18~\kms\ in the plane of the sky has less impact on \WAT\ masers, due
to the much larger velocities of the features (up to $\sim$100~\kms).

In conclusion, our \METH\ measurements can be used (combined with the
corresponding Doppler shifts) to determine the 3D velocity of the star+disk
system. In practice, in order to obtain the velocity in the plane of the sky
{\it with respect to the star}, any absolute proper motion must be corrected
by subtracting {\it the mean absolute proper motion of the \METH\ masers},
i.e. $-16$~\kms\ along the right-ascension axis and +7.6~\kms\ along the
declination.

\section{Disk and jet structure in \I}
\label{sdj}

Better knowledge of the distance and, especially, of the systemic 3D velocity
of \I\ allows us to improve on the model presented by Moscadelli et al.
(\cite{mosca05}). Also, one can accurately compare the distributions and
velocities of the \METH\ masers with those of the \WAT\ masers, and thus draw
a better picture of the disk and jet structure and of their relationship.

Before proceeding to this comparison, it is worth noting that the two types
of masers have been observed at very different epochs, namely November 9 2000
for the \WAT\ masers
(Moscadelli et al.~\cite{mosca05})
and November 6 2004 for the \METH\ masers
(this paper).
The maser spots could have moved over this $\sim$4~yr period, thus questioning a
positional comparison between the two maser types. However, from the known
proper motions one can estimate, e.g., the expected mean position of the
methanol masers at the observing epoch of the water masers, and we find that
the displacement is only $\sim$8~mas in R.A. and $\sim$18~mas in Dec. Values
that small are negligible with respect to the region of interest for us
($\ga$250~mas, see Fig.~\ref{fdj}).

\subsection{The conical \WAT\ jet revised}
\label{scone}

\begin{figure*}
\centering
\resizebox{16cm}{!}{\includegraphics[angle=-90]{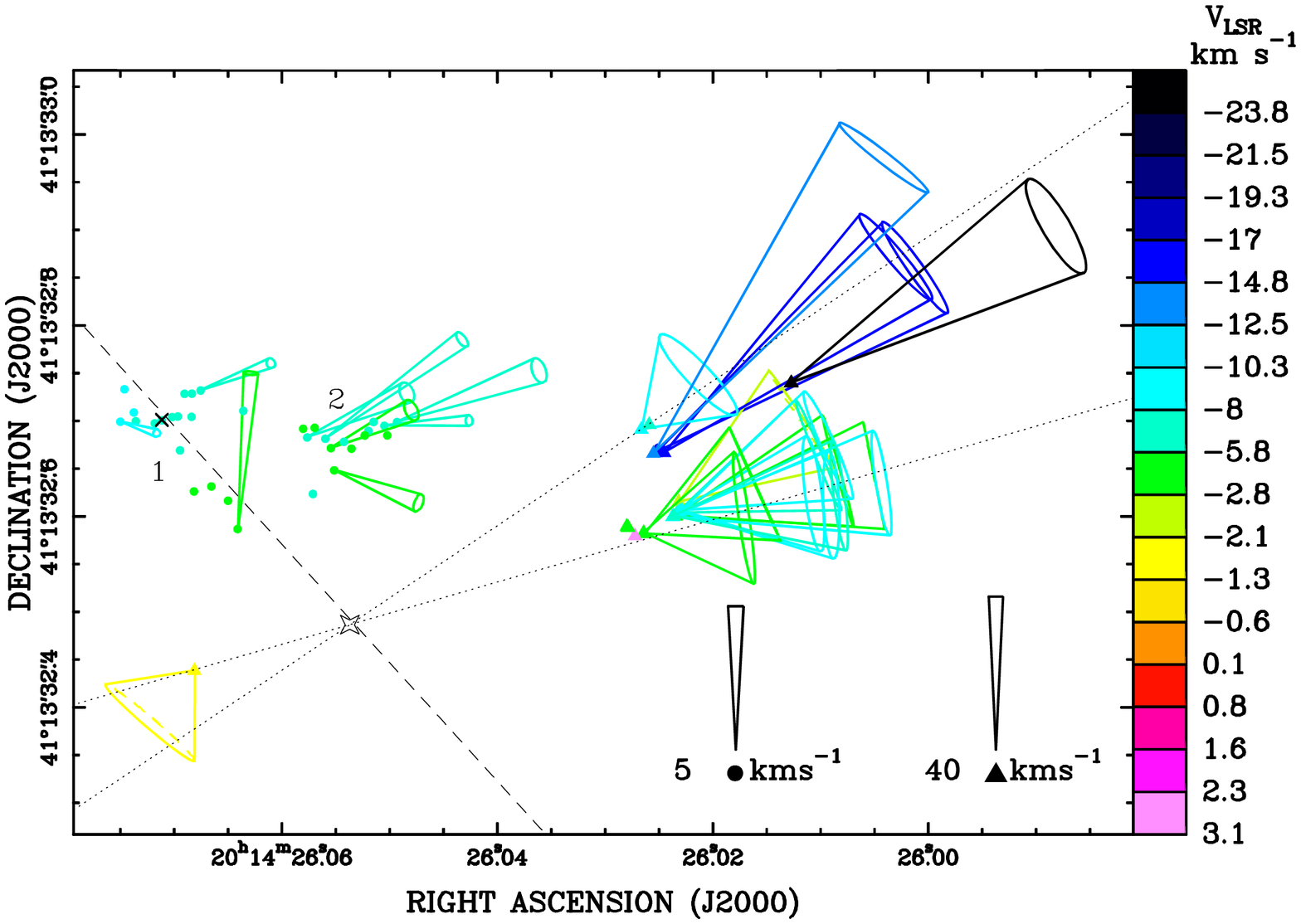}}
\caption{
Water (triangles) and 6.7~GHz methanol (circles) masers in \I. The cones
indicate the 3D velocities of the maser features with respect to the star,
for the \WAT\ masers, or relative to feature~3 (marked with a cross), for the
\METH\ masers.
Points without an associated cone have been detected only over one or
two epochs and the associated proper motion cannot be computed or is considered
unreliable.
The cone opening angle corresponds to the $1\sigma$ uncertainty,
due to the error on the proper motions. The length of the cones
is proportional to the speed, with different scales used for the two maser
types as indicated in the figure. The colours correspond to the line of sight
velocities, as shown in the colour scale to the right. The starred polygon
marks the location of the star, obtained from the model fit to the
\WAT\ maser positions and 3D velocities. Labels ``1'' and ``2'' denote two
elongated groups of \METH\ maser features. The dotted line outline the
projection of the bipolar \WAT\ maser jet onto the plane of the sky, while
the dashed line is a linear fit to the \METH\ maser features of group~1.
}
\label{fdj}
\end{figure*}

Moscadelli et al. (\cite{mosca05}) interpreted the \WAT\ masers as tracing
the surface of a conical jet, expanding away from the star. The model could
satisfactorily fit the observed line-of-sight velocities and absolute proper
motions, after correcting for the parallax, motion of the Sun with respect to
the LSR, and galactic rotation. Knowledge
of the star's 3D velocity, obtained in Sect.~\ref{svsys}, permits
accurate estimation of the \WAT\ maser velocities with respect to the star.
Using this information, we could correct the proper motions of the \WAT\
maser features derived by Moscadelli et al.~(\cite{mosca05}).
The new velocity vectors are shown as arrows in Fig.~\ref{fdj}. Comparison
with Fig.~2 of Moscadelli et al.~(\cite{mosca05}) shows that the correction
does not affect significantly the magnitudes and directions of the vectors,
apart from the red-shifted feature to the SE. This had negligible proper
motion in the old analysis of the data, whereas with the new correction it
appears to move towards SE, lending further support to the bipolar jet model
of the \WAT\ masers.
Note that the proper motions obtained in the present study for the two
features selected for our parallax measurement in Sect.~\ref{spara} 
(--46~\kms in R.A. and 11~\kms in Dec. for the feature with LSR velocity of --3.2~\kms, and
 --47~\kms in R.A. and 17~\kms in Dec. for that with LSR velocity of --6.4~\kms)
are
consistent with the jet-model proposed by Moscadelli et al.~(\cite{mosca05}).
We have not derived the proper motions of the other features detected
in the present experiment, because these will
be the subject of a forthcoming paper illustrating the results of a VLBI
multi-year monitoring of the water maser emission in \I.

We used the model of Moscadelli et al.~(\cite{mosca05}) to fit
the new, corrected data and obtained new best-fit parameters that
are qualitatively
consistent with the old ones. We find a position angle (P.A.) of
115\degr\ instead of 123\degr, an inclination with respect to the line of
sight of 80\degr\ instead of 96\degr, an opening angle of 9\degr\ instead of
17\degr, and a velocity gradient along the jet of 0.116~\kms~mas$^{-1}$
instead of 0.255~\kms~mas$^{-1}$. Also, the position of the star is shifted
by +143~mas in right ascension and --49~mas in declination.

In conclusion, despite some differences between the old and new best-fit
parameters, the corrected data confirm that the jet is very beamed, lies
close to the plane of the sky, and is approximately perpendicular to the disk
(whose P.A. is $\sim53\pm7\degr$; see Cesaroni et al.~\cite{cesa05}).

\subsection{The disk-jet interface}

So far, we have used the \METH\ masers only to determine the systemic
velocity of the star+disk system. Now, we analyse their role in tracing the
distribution and kinematics of the gas in the disk-jet system.

Looking at Fig.~\ref{fdj}, one sees that the \METH\ maser features are
roughly outlining two linear structures, one oriented NE--SW and the other
SE--NW; hereafter we will refer to these as, respectively, ``group~1'' and
``group~2''. The features belonging to these groups are (using the labels in
Table~\ref{tmeth}): 2, 3, 5, 6, 7, 10, 11, 12, 13, 15, 18, 20, 24, 26, 27,
28, 32, 33 for group~1 and 1, 4, 8, 9, 14, 16, 17, 19, 21, 22, 23, 25, 29,
30, 31 for group~2.
Comparison with Edris et al.~(\cite{edris}) lends support to this
distinction, because both the spots positions and
velocities appear consistent with those of our groups~1 and~2, with
only one exception (spot number 15 in their Table~6).

We have calculated the {\it relative} proper motions (see Fig.~\ref{fdj}) of
the \METH\ maser features of both groups with respect to feature~3, which
belongs to group~1 and appears to be the most reliable for the persistency of
the internal structure over the three observing epochs.

It is worth noting two facts. First, the masers in group~2 move towards
NW, i.e. in the same direction as the blue lobe of the \WAT\ maser jet.
Despite the different speeds (up to $\sim$100~\kms for \WAT\ and only
$\sim$5~\kms\ for \METH), this suggests a relationship between the two.
Second, the mean velocity of group~1 features along the line of sight is
--6.9~\kms, which differs by --3.4~\kms\ with respect to the systemic LSR
velocity of --3.5~\kms. The mean distance of the same features from the
nominal position of the star (starred symbol in Fig.~\ref{fdj}) is 0\farcs27
or $\sim$460~AU. If the masers are undergoing Keplerian rotation about the
7~$M_\odot$ star, the expected velocity is $\sim-3.7$~\kms, consistent with
the value of --3.4~\kms\ quoted above.

Based on this, we propose that group~1 masers trace the Keplerian disk (as
proposed by Minier et al.~\cite{mini01} and Edris et al.~\cite{edris}). As
for group~2 masers we speculate that they also lie in the disk, but very
close to the transition region between the disk surface and the outer part of
the jet. Although based on admittedly marginal evidence, this hypothesis may
explain why the relative proper motions of these features (see
Fig.~\ref{fdj}) are roughly perpendicular to the plane of the disk. Besides,
we will show in the following that also the line-of-sight velocities of
group~2 masers
(whose mean value is --5.9~\kms)
differ from those of group~1 masers lending support to our
speculation.
Finally, we note that the number of maser spots without measurable proper
motions is larger in group~2 (40\%) then in group~1 (22\%). This suggests
that masers at the jet interface might be more persistent than
those in the disk plane, consistent with the idea that radiative
excitation of methanol masers may be more effective close to the
surface than inside the densest parts of the disk.

\subsubsection{\METH\ maser group 1: The disk}

\begin{figure}
\centering
\resizebox{8cm}{!}{\includegraphics[angle=-90]{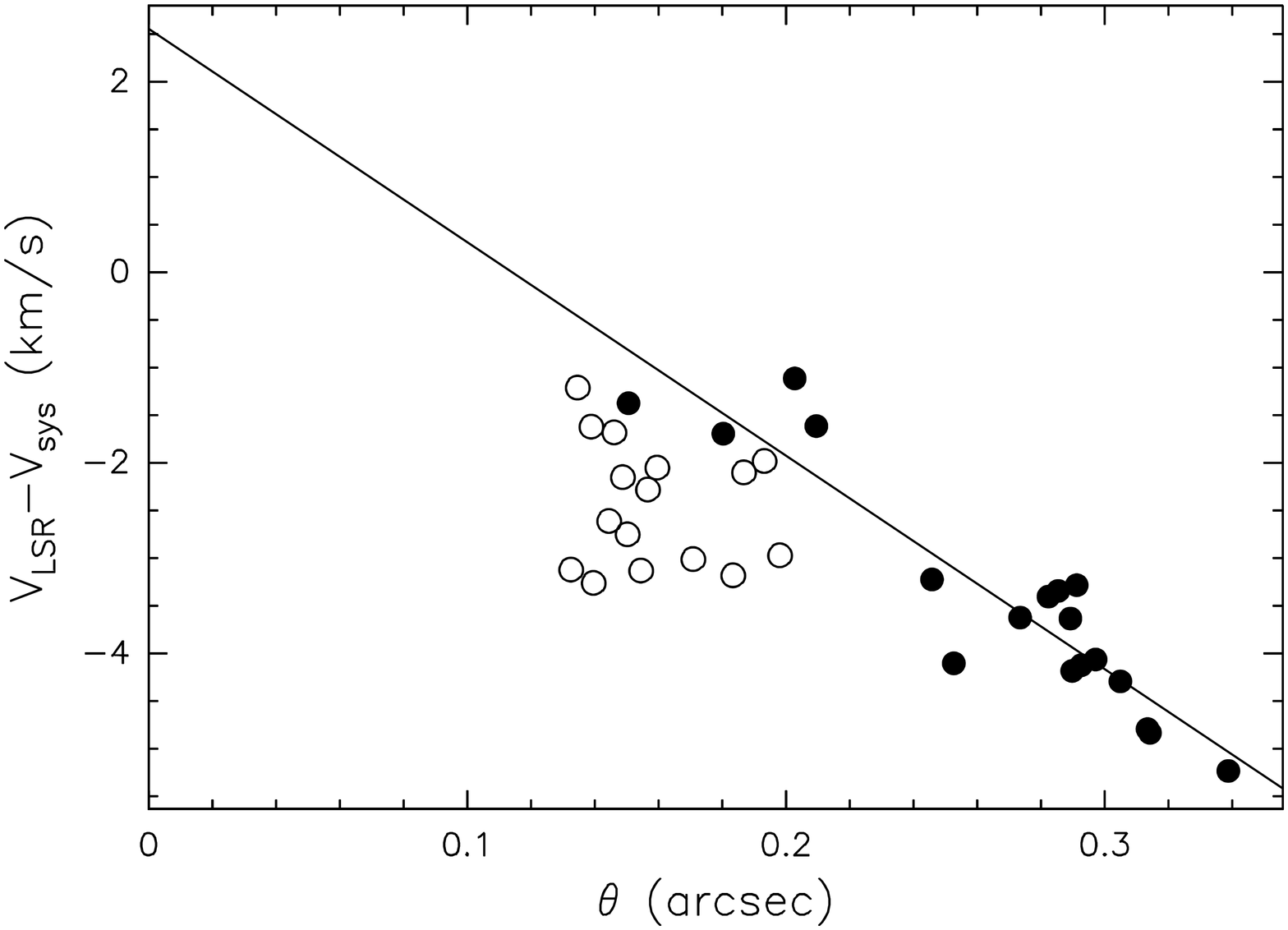}}
\caption{
Plot of the line-of-sight velocity of the \METH\ maser features, relative
to the systemic LSR velocity (--3.5~\kms), versus the angular separation from
the nominal position of the star, measured along the dashed line in
Fig~\ref{fdj}. Solid and empty circles denote, respectively, group~1 and
group~2 features. The line is a linear fit to the points of group~1.
}
\label{fmvr}
\end{figure}

 From Fig.~\ref{fdj} one can see that the nominal position of the star
(obtained from the model fit to the \WAT\ maser jet) lies right along the
dashed line obtained from a linear fit to the features of group~1. This is
consistent with our hypothesis that these masers lie in the plane of the
disk. One can thus study how the component of the rotation velocity along
the line of sight depends on the projected distance from the star. This is
shown in Fig.~\ref{fmvr}, where we plot the difference between the
\METH\ maser velocity and the systemic velocity of \I\ ($\sim-3.5$~\kms)
versus the angular distance, $\theta$, from the nominal position of the star,
measured along the dashed line in Fig.~\ref{fdj}. For the sake of comparison,
we plot also the velocities of group~2 features.

Interestingly, for group~1 the velocity presents a linear dependence
on $\theta$, different from a Keplerian disk where the rotation velocity
should vary with the distance $R$ from the star as $R^{-1/2}$. However, if
the masers are all lying in the disk at the same distance from the star, one
can demonstrate that the {\it line-of-sight} component of the velocity is
proportional to the {\it projected} distance from the star, as in
Fig.~\ref{fmvr}. Why should the masing gas be confined to a narrow ring? One
may speculate that the relevant physical parameters attain the values needed
to trigger the \METH\ population inversion at a specific radius in the disk,
because high densities near the star likely would quench the masers and low
densities far from the star may provide too little gain for strong
amplification.

In this model, the velocity along the line of sight should be zero at the
position of the star, namely at $\theta=0$ in Fig.~\ref{fmvr}. A linear fit
to the solid circles in the figure gives
$(\Vlsr-V_{\rm sys})(\kms)=(-22.4\pm1.1)\,\theta({\rm arcsec})+(2.6\pm0.3)$,
implying that
$(\Vlsr-V_{\rm sys})=0$~\kms\ is attained for $\theta=0\farcs11$. Although
different from zero, this offset is within the uncertainties on the nominal
position of the star (obtained from our model fit to the \WAT\ maser jet).

 From the fit in Fig.~\ref{fmvr} one can also compute a lower limit to the
mass of the star. This is obtained assuming that the rotation speed of the
farthest feature from the star is directed along the line of sight.
Centrifugal equilibrium with $\sim$5~\kms\ at a radius of
0\farcs34--0\farcs11=0\farcs23 (or $\sim$390~AU) implies a mass of
11~$M_\odot$, consistent with the estimate of $7\pm3~M_\odot$ by
Cesaroni et al.~(\cite{cesa05}).

We conclude that the group~1 \METH\ maser features might be
tracing a narrow ring inside the circumstellar disk in \I.
A similar result has been obtained by Torstensson et al.~(\cite{torst},
\cite{torst10}),
who
find that the \METH\ maser spots in Cep~A appear to be all equidistant from
the 7~mm continuum peak. This as well as other similar studies (Bartkiewicz et
al.~\cite{bart05}, \cite{bart09}; Vlemmings et al.~\cite{vlem})
seem to
suggest that not only \METH\ masers form in disks, but also that they are excited at a
specific radius rather than being randomly distributed all over the disk.

\subsubsection{\METH\ maser group 2: The jet}

In Fig.~\ref{fmvr} one can see that all group~2 points fall below the
trend described by group~1 points: as noted above, this appears to
confirm a difference between the origins of the two maser groups.
In particular, we suggest that
the group 2 velocities might be affected by (blue-shifted) expansion along
the jet, which would be entraining part of the material of the disk.
In other words,
we speculate that group~2 features could be tracing the outer regions of the
jet, where the material is lifted from the disk surface and accelerated along
the rotation axis.
This scenario is reminiscent of MHD ``disk winds'' (see,
e.g., Pesenti et al.~\cite{pese} and references therein), invoked to explain
the formation of jets in YSOs. These differ from the alternative model of ``X
winds'' (Shu et al.~\cite{xwin}) basically for two characteristics: the size
of the region from which the wind originates and its speed. While X~winds
predict that the jet should form inside a radius of $\sim$0.02~AU and reach
speeds as large as a few 100~\kms, disk winds should arise from a region
of $\sim$1~AU and be much less biased towards high speeds.

Under the working hypothesis that our interpretation is correct,
one may wonder whether our findings could
be used to discriminate between the two models: disk-wind and X-wind.
In \I, the
group~2 \METH\ masers are moving at a speed of $\sim$5~\kms\ and are located
at a radius of $\sim$300~AU from the disk axis. While these values appear
quite inconsistent with X winds, also a disk-wind interpretation is
doubtful. In fact, 300~AU is by far greater than the radius expected for the
formation of a jet in this type of model and one should invoke a scaled-up
version of disk winds to explain such a large value. However, disk wind
models cannot be simply extrapolated to the high-mass case, because massive
protostars reach the zero-age main sequence still undergoing heavy accretion,
whose impact on jet formation is certainly non negligible. In addition, the
role of magnetic fields relative to gravitational forces may be different in
high-mass star formation with respect to low-mass.

While no safe conclusion can be drawn on the basis of the present data
and even the jet interpretation for group~2 features is questionable,
we believe that further proper motion studies of class~II \METH\ masers
in this source are a unique tool to investigate the structure and velocity
field of the ``twilight zone'' between the disk and the jet, on spatial scales
which cannot be investigated even with ALMA.

\section{Summary and conclusions}

We performed multi-epoch water and 6.7~GHz methanol maser VLBI observations
towards the well known high-mass protostar \I. The \WAT\ maser data are used
to measure the annual parallax of the source and thus obtain an accurate
distance of $1.64\pm0.05$~kpc. This is consistent
with the value adopted so far in the literature (1.7~kpc) and confirms that
\I\ is the best example of an early B-type protostar associated with a
Keplerian disk and bipolar jet.

We also derived the absolute proper motions of the \METH\ masers, which
gives the 3D velocity of the disk+star system. This can be used to correct the
\WAT\ maser velocities given by Moscadelli et al.~(\cite{mosca05}) and
optimize their model fit. The best-fit parameters are qualitatively consistent
with those obtained by Moscadelli et al.~(\cite{mosca05}), confirming that the
jet lies very close to the plane of the sky and is roughly perpendicular to
the disk.

By referring the proper motions of the \METH\ masers to a feature lying in
the disk plane and analysing their distribution and kinematics, we identify
two groups of maser features: one associated with the disk, the other possibly
tracing the disk at the interface with the bipolar jet. We speculate that
class~II \METH\ masers are an ideal tracer to investigate the dense
molecular gas entrained at the basis of the jet.

\begin{acknowledgements}
 We thank Francesca Bacciotti for useful discussions about the properties
 of X-winds and disk-winds. We also thank the referee, Gary Fuller, for his
 careful reading of the manuscript and useful suggestions. This work has
 benefited from research funding from the European Community's sixth
 Framework Programme under RadioNet R113CT 2003 5058187 and has been partly
 supported by the European Community Framework Programme 7, Advanced Radio
 Astronomy in Europe, grant agreement No. 227290.
\end{acknowledgements}


\begin{thebibliography}{}

%\bibitem[1996]{angla}
% Anglada, G. 1996, ASP Conference Series, 93, 3 
%\bibitem[2008]{araya}
% Araya, E., Hofner, P., Kurtz, S., Olmi, L., \& Linz, H. 2008, ApJ, 675, 420
\bibitem[2005]{bart05}
 Bartkiewicz, A., Szymczak, M., \& van Langevelde, H. J. 2005, A\&A, 442, L61
\bibitem[2009]{bart09}
 Bartkiewicz, A., Szymczak, M., van Langevelde, H. J., Richards, A. M. S.,
 \& Pihlstr\"om, Y. M. 2009, A\&A, 502, 155
%\bibitem[2004]{bel04}
% Beltr\'an, M.T, Cesaroni, R., Neri, R., et al. 2004, ApJ, 601, L187
%\bibitem[2005]{bel05}
% Beltr\'an, M.T, Cesaroni, R., Neri, R., et al. 2005, A\&A, 435, 901
%\bibitem[2006]{bel06}
% Beltr\'an, M.T, Cesaroni, R., Codella, C., et al. 2006, Nature, 443, 427
%\bibitem[2007]{beuth}
% Beuther, H., Churchwell, E.B., McKee, C.F., \& Tan, J.C. 2007, in
% Protostars and Planets V, ed. B. Reipurth, D. Jewitt, \& K. Keil
% (Tucson: Univ. of Arizona Press), 165
%\bibitem[2005]{iaus}
% Cesaroni, R. 2005, in Massive Star Birth: a Crossroads of Astrophysics, IAU
% Symposium 227, ed. R. Cesaroni, M. Felli, E. Churchwell, M. Walmsley
% (Cambridge University Press), 59
%\bibitem[1994a]{cesa94}
% Cesaroni, R., Churchwell, E., Hofner, P., Walmsley, C.M., \& Kurtz, S. 1994a,
% A\&A, 288, 903
%\bibitem[1994b]{cesag31}
% Cesaroni, R., Olmi, L., Walmsley, C.M., Churchwell, E., \& Hofner, P. 1994b,
% ApJ, 435, L137
%\bibitem[1998]{cesa98}
% Cesaroni, R., Hofner, P., Walmsley, C.M., \& Churchwell, E. 1998, A\&A,
% 331, 709
%\bibitem[1990]{chur90}
%Churchwell E., Walmsley C.M., \& Cesaroni R. 1990, A\&AS, 83, 119
%\bibitem[1990]{fran}
% Franco, J., Tenorio-Tagle, G., \& Bodenheimer, P. 1990, ApJ, 349, 126
%\bibitem[2005]{furu05}
% Furuya, R. S., Cesaroni, R., Takahashi, S. et al. 2005, ApJ, 624, 827
%\bibitem[2002]{fonta02}
% Fontani, F., Cesaroni, R., Caselli, P., \& Olmi, L. 2002, A\&A, 389, 603
%\bibitem[2004]{gibb}
% Gibb, A.G., Wyrowski, F., \& Mundy, L.G. 2004, ApJ, 616, 301
%\bibitem[2009]{gira}
% Girart, J.M., Beltr\'an, M.T., Zhang, Q., Rao, \& R., Estalella, R. 2009,
% Science, 324, 1408
%\bibitem[2007]{hoare}
% Hoare, M.G., Kurtz, S., Lizano, S., Keto, E., \& Hofner, P. 2007, in
% Protostars and Planets V, ed. B. Reipurth, D. Jewitt, \& K. Keil
% (Tucson: Univ. of Arizona Press), 181
%\bibitem[2002]{keto02}
% Keto, E.H. 2002, ApJ, 580, 980
%\bibitem[2003]{keto03}
% Keto, E.H. 2003, ApJ, 599, 1196
%\bibitem[2007]{keto07}
% Keto, E.H. 2007, ApJ, 666, 976
%\bibitem[2006]{kewo}
% Keto, E.H. \& Wood, K. 2006, ApJ, 637, 850
%\bibitem[2009]{lopsep}
% L\'opez-Sepulcre, A., Codella, C., Cesaroni, R., Marcelino, N.,
% \& Walmsley, C.M. 2009, A\&A, 499, 811
%\bibitem[2002]{mart}
% Martins, F., Schaerer, D., Hillier, D.J. 2002, A\&A 382, 999
%\bibitem[2001]{maxia}
% Maxia, C., Testi, L., Cesaroni, R., \& Walmsley, C.M. 2001, A\&A 371, 287
%\bibitem[1996]{myers}
% Myers, P.C., Mardones, D., Tafalla, M., Williams, J.P., \& Wilner, D.J. 1996,
% ApJ, 465, L133
%\bibitem[1996a]{olmi96a}
% Olmi, L., Cesaroni, R., \& Walmsley, C.M. 1996a, A\&A 307, 599
%\bibitem[1996b]{olmi96b}
% Olmi, L., Cesaroni, R., Neri, R., \& Walmsley, C.M. 1996b, A\&A 315, 565
%\bibitem[2009]{osor}
% Osorio, M., Anglada, G., Lizano, S., \& D'Alessio, P. 2009, ApJ 694, 29
%\bibitem[1973]{pana}
% Panagia, N. 1973, AJ, 78, 929
%\bibitem[1975]{pafe}
% Panagia, N. \& Felli, M. 1975, A\&A, 39, 1
%\bibitem[2008]{pand}
% Pandian, J.D., Momjian, E., \& Goldsmith, P.F. 2008, A\&A, 486, 191
%\bibitem[1986]{unge}
% Ungerechts, H., Walmsley, C.M., \& Winnewisser, G. 1986, A\&A, 157, 207
%\bibitem[1993]{vand}
% Van Dishoeck, E.F., Blake, G.A., Draine, B.T., \& Lunine, J.I. 1993,
% Protostars and Planets III, ed. E.H. Levy \& J.I. Lunine,
% (Tucson: Univ. of Arizona Press), 163
%\bibitem[2001]{wiln}
% Wilner, D.J., De Pree, C.G., Welch, W.J., \& Goss, W.M. 2001, ApJ, 550, L81
%\bibitem[1989]{wca}
% Wood, D.O.S. \& Churchwell, E. 1989, ApJ 69, 831
%\bibitem[2004]{wu04}
% Wu, Y., Wei, Y., \& Zhao, M., 2004, A\&A, 426, 503
%\bibitem[1985]{yorke}
%Yorke, H.W. 1985, in Birth and Infancy of Stars, p. 645
%\bibitem[2005]{zhang}
% Zhang, Q. 2005, in Massive Star Birth: a Crossroads of Astrophysics, IAU
% Symposium 227, ed. R. Cesaroni, M. Felli, E. Churchwell, M. Walmsley
% (Cambridge University Press), 135
\bibitem[2008]{caga}
 Caratti o Garatti, A., Froebrich, D., Eisl{\"o}ffel, J., Giannini, T., 
 \& Nisini, B.\ 2008, A\&A, 485, 137 
\bibitem[1997]{cesa97}
 Cesaroni, R., Felli, M., Testi, L., Walmsley, C.~M., 
 \& Olmi, L.\ 1997, A\&A, 325, 725 
\bibitem[1999]{cesa99}
 Cesaroni, R., Felli, M., Jenness, T., Neri, R., Olmi, L., Robberto, M., 
 Testi, L., \& Walmsley, C.~M.\ 1999, A\&A, 345, 949 
\bibitem[2005]{cesa05}
 Cesaroni, R., Neri, R., Olmi, L., Testi, L., Walmsley, C.~M., 
 \& Hofner, P.\ 2005, A\&A, 434, 1039 
\bibitem[2006]{natu}
 Cesaroni R., Galli D., Lodato G., Walmsley M., Zhang Q. 2006, Nature 444, 703
\bibitem[2007]{ppv}
 Cesaroni, R., Galli, D., Lodato, G., Walmsley, C.M., \& Zhang, Q. 2007, in
 Protostars and Planets V, ed. B. Reipurth, D. Jewitt, \& K. Keil
 (Tucson: Univ. of Arizona Press), 197
\bibitem[2007]{debu}
 De Buizer, J. M.\ 2007, ApJ, 654, L147
\bibitem[2005]{edris}
 Edris, K.~A., Fuller, G.~A., Cohen, R.~J., \& Etoka, S.\ 2005, A\&A, 434, 213 
\bibitem[2007]{felli}
 Felli, M., Brand, J., Cesaroni, R., et al. 2007, A\&A, 476, 373 
\bibitem[1999]{hof99}
 Hofner, P., Cesaroni, R., Rodr{\'{\i}}guez, L.~F., \& Mart{\'{\i}}, J.\ 1999,
 A\&A, 345, L43 
\bibitem[2007]{hof07}
 Hofner, P., Cesaroni, R., Olmi, L., Rodr{\'{\i}}guez, L.~F., Mart{\'{\i}}, 
 J., \& Araya, E.\ 2007, A\&A, 465, 197 
\bibitem[2009]{krum}
 Krumholz, M.R., Klein, R.I., McKee, C.F., Offner, S.S.R., \& Cunningham, A.J.
 2009, Science, 323, 754
\bibitem[2006]{lebr}
 Lebr{\'o}n, M., Beuther, H., Schilke, P., \& Stanke, Th.\ 2006, A\&A, 448, 1037
\bibitem[1981]{mibi}
 Mihalas, D. \& Binney, J. 1981, Galactic Astronomy
 (San Francisco: W.H. Freeman and Company)
\bibitem[2000]{mini00}
 Minier, V., Booth, R.~S., \& Conway, J.~E.\ 2000, A\&A, 362, 1093 
\bibitem[2001]{mini01}
 Minier, V., Conway, J.~E., \& Booth, R.~S.\ 2001, A\&A, 369, 278 
\bibitem[2000]{mosca00}
 Moscadelli, L., Cesaroni, R., \& Rioja, M.~J.\ 2000, A\&A, 360, 663 
\bibitem[2005]{mosca05}
 Moscadelli, L., Cesaroni, R., \& Rioja, M.~J.\ 2005, A\&A, 438, 889 
\bibitem[1998]{norris}
 Norris, R.~P., Byleveld, S.~E., Diamond, P.~J., et al. 1998, ApJ, 508, 275 
 %Norris, R.~P., Byleveld, S.~E., Diamond, P.~J., Ellingsen, S.~P., Ferris, 
 %R.~H., Gough, R.~G., Kesteven, M.~J., McCulloch, P.~M., Phillips, C.~J., 
 %Reynolds, J.~E., Tzioumis, A.~K., Takahashi, Y., Troup, E.~R., 
 %\& Wellington, K.~J.\ 1998, ApJ, 508, 275 
\bibitem[1993]{past}
 Palla, F. \& Stahler, S.W. 1993, ApJ, 418, 414
\bibitem[2004]{pese}
 Pesenti, N., Dougados, C., Cabrit, S., et al. 2004, A\&A, 416, L9
\bibitem[2009]{pesta}
 Pestalozzi, M.~R., Elitzur, M., \& Conway, J.~E.\ 2009, A\&A, 501, 999 
\bibitem[2009b]{reid09b}
 Reid, M.~J., Menten, K.~M., Zheng, X.~W., et al. 2009b, ApJ, 700, 137
 %Reid, M.~J., Menten, K.~M., Zheng, X.~W., Brunthaler, A., Moscadelli, L., 
 %Xu, Y., Zhang, B., Sato, M., Honma, M., Hirota, T., Hachisuka, K., Choi, 
 %Y.~K., Moellenbrock, G.~A., \& Bartkiewicz, A.\ 2009b, ApJ, 700, 137
\bibitem[2009a]{reid09a}
 Reid, M.~J., Menten, K.~M., Brunthaler, A., Zheng, X.~W., 
Moscadelli, L., Xu, Y.\ 2009a, ApJ, 693, 397
\bibitem[2010]{sanna}
Sanna, A., Moscadelli, L., Cesaroni, R., Tarchi, A., 
Furuya, R. S., Goddi, C. 2010, A\&A, in press
\bibitem[2000]{shep}
 Shepherd, D.~S., Yu, K.~C., Bally, J., \& Testi, L.\ 2000, ApJ, 535, 833 
\bibitem[2008]{shin}
 Shinnaga, H., Phillips, T. G., Furuya, R. S., 
 \& Cesaroni, R. 2008, ApJ, 682, 1103 
\bibitem[2000]{xwin}
 Shu, F.H., Najita, J.R., Shang, H., Li, Z.-Y. 2000, in
 Protostars and Planets IV, ed. V. Mannings, A. Boss, S. Russel
 (Tucson: Univ. of Arizona Press), 789
\bibitem[2005]{srid}
 Sridharan, T.~K., Williams, S.~J., \& Fuller, G.~A.\ 2005, ApJ, 631, L73 
\bibitem[2007]{su}
 Su, Y.-N., Liu, S.-Y., Chen, H.-R., Zhang, Q., \& Cesaroni, R. 2007, ApJ, 671, 571 
\bibitem[2008]{torst}
 Torstensson, K., van Langevelde, H. J., Vlemmings, W., \& van der Tak, F. 2008,
 9th EVN Symposium, p. 39
\bibitem[2010]{torst10}
 Torstensson, K., van Langevelde, H. J., Vlemmings, W., \& Bourke, S. 2010,
 A\&A, in press
\bibitem[2005]{trini}
 Trinidad, M. A., Curiel, S., Migenes, V., et al. 2005, AJ, 130, 2206
% Trinidad, M. A., Curiel, S., Migenes, V., Patel, N., 
% Torrelles, J. M., G{\'o}mez, J. F., Rodr{\'{\i}}guez, L. F., 
% Ho, P. T.~P., \& Cant{\'o}, J. 2005, AJ, 130, 2206
\bibitem[2010]{vlem}
 Vlemmings, W. H. T., Surcis, G., Torstensson, K. J. E., \& van Langevelde, H. J.
 2010, MNRAS, 404, 134
\bibitem[1990]{wilk}
 Wilking, B. A., Blackwell, J. H., Mundy, L. G. 1990, AJ, 100, 758
\bibitem[1998]{zhang}
 Zhang, Q., Hunter, T. R., \& Sridharan, T.~K.\ 1998, ApJ, 505, L151

\end{thebibliography}
\end{document}